\documentclass[sigconf, 10pt, nonacm]{acmart}
\usepackage{blindtext}
\usepackage[utf8]{inputenc}
\usepackage{epsfig,xspace,url,xcolor}
\usepackage{subcaption}
\usepackage{amsmath}
\usepackage{mathtools}
\usepackage{amsthm}
\usepackage{tikz}
\usepackage{caption}
\usepackage[small,compact]{titlesec}
\titlespacing*{\section}{0pt}{4pt}{4pt}
\titlespacing*{\subsection}{0pt}{4pt}{4pt}
\titlespacing*{\subsubsection}{0pt}{0pt}{0pt}

\usepackage{graphicx}
\usepackage[ruled, lined, linesnumbered, commentsnumbered, longend]{algorithm2e}
\usepackage[noend]{algpseudocode}
\usepackage{url}
\usepackage{colortbl}
\usepackage{thmtools}
\usepackage{thm-restate}

\usepackage{wrapfig}
\usepackage[subtle]{savetrees}
\usepackage[framemethod=TikZ]{mdframed}
\renewcommand\footnotetextcopyrightpermission[1]{} 
\input{macros}

\let\ACMmaketitle=\maketitle
\renewcommand{\maketitle}{\begingroup\let\footnote=\thanks \ACMmaketitle\endgroup}
\let\height\relax

\begin{document}

\sloppy
\newcommand{\name}{\textsc{STyGIANet}\xspace}

\title{When Light Bends to the Collective Will: A Theory and Vision for Adaptive Photonic Scale-up Domains}

\author{Vamsi Addanki}
\affiliation{
  \institution{Purdue University}
  \country{USA}
}

\renewcommand{\shortauthors}{Addanki et al.}

\begin{abstract}
As chip-to-chip silicon photonics gain traction for their bandwidth and energy efficiency, collective communication has emerged as a critical bottleneck in scale-up systems. Programmable photonic interconnects offer a promising path forward: by dynamically reconfiguring the fabric, they can establish direct, high-bandwidth optical paths between communicating endpoints --- \emph{synchronously and guided by the structure of collective operations} (e.g., AllReduce). However, realizing this vision --- \emph{when light bends to the collective will} --- requires navigating a fundamental trade-off between reconfiguration delay and the performance gains of adaptive topologies.

In this paper, we present a simple theoretical framework for adaptive photonic scale-up domains that makes this trade-off explicit and clarifies when reconfiguration is worthwhile. Along the way, we highlight a connection --- not surprising but still powerful --- between the Birkhoff--von Neumann (BvN) decomposition, maximum concurrent flow (a classic measure of network throughput), and the well-known $\alpha$--$\beta$ cost model for collectives. Finally, we outline a research agenda in algorithm design and systems integration that can build on this foundation.
\end{abstract}

\maketitle
\thispagestyle{plain}
\pagestyle{plain}

\section{Introduction}
\label{sec:introduction}

The massive scale of modern distributed computing~\cite{NEURIPS2020_1457c0d6,shoeybi2019megatron,10.1145/3458817.3476209,10.1145/3579371.3589350,295551,10.1145/3651890.3672265,10.1145/3651890.3672233} --- spanning hyperscale data centers to tightly-coupled HPC clusters --- has made efficient collective communication a critical bottleneck~\cite{10664412,10.1145/3663408.3663409,9912504}. Scale-up\footnote{``Scale-up'' refers to networks within a single server or a single memory domain.} networks typically connect multiple GPUs using high-bandwidth electrical links, often through electrical switches (e.g., NVSwitches) or PCIe memory interconnects~\cite{nvidia2023superpod}. While these designs have served well for decades, they now face fundamental limits: the bandwidth demands of modern AI workloads and the sheer scale of distributed systems are pushing electrical interconnects beyond what they can practically deliver~\cite{10477550}. As the number of GPUs grows, these links become a bottleneck, creating a bandwidth wall alongside rising power consumption and heat dissipation~\cite{10946778}. The slowdown of Moore's Law for CMOS only compounds this challenge~\cite{10.1145/3387514.3406221,10.1145/3651890.3672273}. As scale-up systems grow larger and more heterogeneous, the need for fundamentally more efficient, scalable, and low-power communication fabrics has never been greater.

In this context, silicon photonics offers a promising path forward, delivering orders-of-magnitude improvements in bandwidth density and energy efficiency compared to electrical interconnects~\cite{10.1145/3696348.3696856,9007742}. By using optical signals for chip-to-chip communication, silicon photonics can dramatically boost data throughput while reducing power draw. Yet despite its promise, much of this potential remains untapped in scale-up systems---primarily because current designs are rigid, lacking the adaptability needed to match dynamic workload patterns such as collective communication.

Photonic interconnects have traditionally been built as static circuit-switched topologies, tuned for specific, often predictable, communication patterns. But this rigidity is starting to crack: programmable silicon photonic fabrics are emerging~\cite{Ding:25}, enabling dynamic reconfiguration of optical paths to adapt to shifting workload demands. These fabrics can establish direct, high-bandwidth optical links between endpoints, unlocking more efficient data exchange and synchronization across GPUs within a scale-up domain.  
Recent work has shown how to schedule circuit-switch configurations that align with communication patterns, using Birkhoff--von Neumann (BvN) decompositions or by solving optimization problems on the aggregate demand matrix~\cite{10.1145/3452296.3472900,285119,10.1145/2716281.2836126,10.1145/1851182.1851223,10.1145/2486001.2486007}. Yet despite this progress, we know surprisingly little about how reconfiguration delays shape collective performance --- or when it is worth reconfiguring at all.

In this paper, we make two simple yet striking observations. First, many collective communication algorithms naturally induce BvN decompositions: each algorithm step can be seen as a matching, and together these matchings form a convex combination of the aggregate demand. This connection has hovered in the literature for years, thanks to the inherently point-to-point, step-wise design of collective algorithms~\cite{https://doi.org/10.1002/cpe.1206}. Second, this perspective bridges neatly to performance modeling: the classic $\alpha$--$\beta$ cost model for collectives emerges naturally when each step is viewed through the lens of maximum concurrent flow, capturing both network throughput and congestion. Together, these insights ground the familiar $\alpha$--$\beta$ model in physical topologies, generalizing collective completion time to account for real-world network constraints.

Leveraging these insights, we present a theoretical framework for optimizing circuit switching in adaptive photonic interconnects. Our focus is the fundamental trade-off between reconfiguration delay and the performance gains enabled by dynamically matching the topology to the communication pattern. We formulate an optimization problem that captures this trade-off, allowing us to systematically decide when to reconfigure the interconnect and when to maintain a static topology, with the objective of minimizing collective completion time. This framework provides a principled way to design circuit-switching schedules that balance the benefits of reconfiguration against its costs, while explicitly accounting for both network throughput and the structure of collective communication.

Our preliminary results show that adaptive circuit switching can unlock substantial performance gains for collective communication --- but only when used wisely. In regimes with high reconfiguration delays or small messages, naive per-step reconfiguration can add more latency than it saves; here, our framework shows when it is better to stay static. Conversely, when delays are low and message sizes are large, carefully chosen reconfigurations can fully tap the available photonic bandwidth, outperforming static designs by a wide margin. Most importantly, we uncover a practical middle ground where neither always reconfiguring nor always staying static is sufficient: this regime clearly requires optimized schedules that decide when reconfiguration is worth the cost and when it is not.

These results expose rich new questions at the intersection of theory and practice: how to design fast heuristics, develop collective algorithms that are reconfiguration-aware, and build photonic fabrics that can adapt on the fly. We outline a research agenda addressing these challenges at the end of this paper. We believe this line of work pushes us closer to interconnects where light truly bends to the collective will.

\section{Background \& Motivation}
\label{sec:bottlenecks}

Collective operations, such as \texttt{AllReduce} and \texttt{Broadcast}, are foundational in distributed computing~\cite{doi:10.1177/1094342005051521,https://doi.org/10.1002/cpe.1206}. These operations progress in structured stages with predictable communication patterns and data dependencies. Yet static interconnects --- even when combined with topology-aware algorithms --- often fail to fully leverage this structure.

\myitem{Limits of topology-aware collectives:}
Prior work has developed collective algorithms tailored for specific static topologies (e.g., torus, DGX), guided by the classic $\alpha$--$\beta$ cost model~\cite{10.1145/3437801.3441620,285084,10.1145/3651890.3672249}. While these designs improve efficiency for fixed topologies, they fundamentally inherit the rigidity of static networks. For example, multi-step collectives like halving/doubling for \texttt{AllReduce}~\cite{10.1007/978-3-540-24685-5_1} require repeated pairwise exchanges, but a fixed topology forces some pairs to traverse longer or congested paths, increasing both latency and bandwidth requirements~\cite{295653}. Worse, static networks must provision for worst-case demand, leading to underutilization when traffic is sparse or staged. This is often tackled by pipelining and mirroring the collectives for multi-ported networks~\cite{295653}, but this approach only partially mitigates the inefficiencies of static designs.

\myitem{Throughput modeling and BvN decompositions:}
The maximum concurrent flow~\cite{10.1145/77600.77620} framework has long been used to analyze network throughput and congestion~\cite{jyothi2016measuring,highthroughputSingla,10.1145/3452296.3472913,10.1145/3579312,10.1145/3519935.3520020}. It connects naturally to Birkhoff--von Neumann (BvN) decompositions, which express an aggregate traffic matrix as a convex combination of matchings~\cite{10.1145/3452296.3472913}. Many works use an aggregate traffic matrix as an input to synthesize circuit-switching schedules for demand-aware networks~\cite{10.1145/2486001.2486007,285119,10.1145/3452296.3472900}. Yet, traffic matrices, and BvN decompositions assume all traffic is available simultaneously --- which is not true for collectives that generate and exchange data in a strict sequence. This mismatch means static traffic matrix decompositions alone cannot capture the temporal dependencies that real collectives impose.

\myitem{Programmable but costly reconfiguration:}  
Reconfigurable photonic fabrics hold the promise of tailoring the network topology to each step of a collective communication pattern, reducing congestion and boosting throughput~\cite{Ding:25,10.1145/3696348.3696856}. But this flexibility comes at a price: practical designs introduce non-trivial reconfiguration delays, which can easily wipe out any performance gains if applied without care~\cite{Ding:25}. Yet much of the existing work sidesteps this trade-off altogether, either assuming that reconfiguration overheads are negligible or simply falling back to static networks when they are not.

Together, these gaps motivate a more principled perspective — one that bridges the staged structure of collective algorithms, the limits of network throughput, and the real costs of reconfiguration. In the next section, we present a framework that connects BvN decompositions, maximum concurrent flow, and the $\alpha$--$\beta$ model, providing fresh insight into when and how adaptive photonic interconnects can truly pay off.

\section{Theory for Adaptive Scaleup Domains}

We present a theoretical framework for optimizing circuit switching in adaptive photonic interconnects, focusing on the trade-off between reconfiguration delay and performance gains of adaptive topologies. 

\subsection{Architecture and Assumptions}
\label{sec:architecture}

We consider a scale-up domain with $n$ GPUS, each equipped with an electrical-to-optical tranceiver (e.g., TeraPhy~\cite{9007742}) with bandwidth $b$. All the $n$ tranceivers are connected to a photonic interconnect with $n$ ports. Light enters through these ports and can be routed through the interconnect that establishes direct optical paths between pairs of ports --- essentially connecting two GPUs. The interconnect is programmable i.e., reconfigurable, allowing it to dynamically reconfigure the optical paths on-demand~\cite{10.1145/3696348.3696856}. Alternatively, if the tranceivers are capable of tuning the wavelength of the light they emit, a passive wavelength switching photonic interconnect can establish direct paths between pairs of ports, without requiring a cental controller. In either designs, we consider a reconfiguration delay of $\alpha_r$ for the interconnect to reconfigure the optical paths. We note that several technologies today incur a reconfiguration delay that is dependent on the number of ports involved in the reconfiguration~\cite{Ding:25}. For simplicity, we assume the reconfiguration delay $\alpha_r$ is constant for all reconfigurations (e.g., for the total port count), but our framework can be extended to account for this variability. Importantly, we assume that all GPUs are within a single scale-up domain, and thus have fast access to a shared memory (e.g., DGX H100 server~\cite{9895480}). This allows the GPUs to rapidly synchronize e.g., using a barrier, before a particular step during a collective, so that they can perform the reconfiguration (if required) synchronously and proceed to the next step. We currently focus on collective communication across all $n$ GPUs. A subset of GPUs can also be considered, and the interconnect simply reconfigures (if required) only the involved ports.

\subsection{BvN, Concurrent Flow, and the $\alpha$--$\beta$ Cost Model}
\label{sec:concurrent-flow}

We model collective communication performed by an algorithm across $n$ GPUs as a sequence of $s$ communication steps. In each step $i$, a fixed amount of data $m_i$ is exchanged between pairs of GPUs according to a matching, represented by a permutation matrix $\mathcal{M}_i$. Each entry $\mathcal{M}_i(j, k) = 1$ indicates that GPU $j$ sends data to GPU $k$ during step $i$; all other entries are zero. The full collective communication algorithm can thus be described as a sequence $\langle \mathcal{M}_1, \mathcal{M}_2, \ldots, \mathcal{M}_s \rangle$ of permutations, with associated data volumes $\langle m_1, m_2, \ldots, m_s \rangle$.

The total communication across all steps can be captured by the \emph{aggregate demand matrix} $\mathcal{M}$, where each entry $\mathcal{M}(j, k)$ denotes the total volume of data sent from GPU $j$ to GPU $k$ over the entire operation. This matrix is simply the sum of all stepwise permutation matrices, weighted by their data volumes:
\begin{equation}
\mathcal{M} = m_1 \cdot \mathcal{M}_1 + m_2 \cdot \mathcal{M}_2 + \ldots + m_s \cdot \mathcal{M}_s.
\end{equation}

This expression is, by definition, a \emph{Birkhoff--von Neumann (BvN) decomposition} of $\mathcal{M}$: a convex combination of permutation matrices. In this view, the steps of the collective algorithm naturally correspond to matchings in the decomposition, with each $m_i$ denoting the volume of data transferred during step $i$.

\begin{graybox}
\begin{observation}[Collectives Induce BvN Decompositions]
Collective communication algorithms that proceed via a sequence of matchings naturally induce a BvN decomposition of their aggregate demand matrix.
\end{observation}
\end{graybox}

The reverse, however, does not hold: not all BvN decompositions correspond to valid collective algorithms. More critically, BvN decompositions fail to capture the \emph{temporal structure} inherent in collective communication. In real algorithms, the ordering of permutations matters—steps cannot be arbitrarily rearranged. The data exchanged in step $i$ is often \emph{generated as a result of} the computation or communication in step $i-1$, creating a strict sequence of dependencies.

These temporal and data-flow constraints underscore an important limitation: the aggregate demand matrix, while useful in demand-aware network design~\cite{10.1145/2934872.2934911,10.1145/2716281.2836126,10.1145/2486001.2486007}, assumes all traffic is simultaneously available between source-destination pairs. This assumption breaks down in collectives, where data availability is staged and communication steps must follow a strict temporal order. As a result, designing interconnects for collective communication requires reasoning beyond static demand matrices and BvN decompositions alone.

Yet, the BvN decompositions induced by collective algorithms, as we show next, reveal a useful connection to both network throughput and the classic $\alpha$–$\beta$ cost model.

Consider a graph $G = (V, E)$, where $V$ is the set of $n$ GPUs and $E$ represents the photonic links between them. The total completion time of the collective communication algorithm can be expressed as:
\begin{align}
t_c & = DCT(m_1 \cdot \mathcal{M}_1) + DCT(m_2 \cdot \mathcal{M}_2) + \ldots + DCT(m_s \cdot \mathcal{M}_s),
\end{align}
where $DCT(m_i \cdot \mathcal{M}_i)$ denotes the \emph{demand completion time} of step $i$, corresponding to a data volume $m_i$ and communication pattern $\mathcal{M}_i$.

The value of $DCT(m_i \cdot \mathcal{M}_i)$ depends on the structure and capacity of the underlying graph $G$. Specifically, we define the \emph{maximum concurrent flow} $\theta(G, \mathcal{M}_i)$ as the largest fraction of the permutation demand matrix $\mathcal{M}_i$ that can be routed simultaneously without exceeding any link capacities. Intuitively, $\theta(G, \mathcal{M}_i)$ quantifies the achievable throughput for that step’s communication pattern. This implies that the demand completion time can be written as:
\[
DCT(m_i \cdot \mathcal{M}_i) = \frac{m_i}{b} \cdot \frac{1}{\theta(G, \mathcal{M}_i)},
\]
where $b$ is the link bandwidth. Here, $\frac{m_i}{b}$ represents the ideal transmission time assuming full throughput, while the factor $\frac{1}{\theta(G, \mathcal{M}_i)}$ accounts for congestion. By definition of the maximum concurrent flow, the effective bandwidth available for this permutation is $b \cdot \theta(G, \mathcal{M}_i)$, so the actual transmission time scales inversely with the achievable throughput.

In addition, each communication step $i$ incurs a fixed overhead $\alpha$, which captures startup latencies such as data preparation; latency $\delta \cdot \ell_i$ incurred due to per-link propagation delay $\delta$ and the path length $\ell_i$ of the most congested link in the corresponding step, which is often neglected and absorbed into the constant $\alpha$.
If the network offers bandwidth $b$ per node, we define $\beta = \frac{1}{b}$. The demand completion time for step~$i$ can then be written as:
\begin{align}
DCT(m_i \cdot \mathcal{M}_i) =  
\underbrace{\alpha + \delta \cdot \ell_i}_{\text{latency\ factor}} + 
\overbrace{\beta}^{\text{bandwidth\ factor}} \cdot m_i \cdot 
\underbrace{\frac{1}{\theta(G, \mathcal{M}_i)}}_{\text{congestion\ factor}}
\end{align}

\noindent 
The total completion time of the collective for all the $s$ steps can now be expressed as:
\begin{align}
t_c &= \sum_{i=1}^{s} DCT(m_i \cdot \mathcal{M}_i) = \sum_{i=1}^{s} \left( \alpha + \delta\cdot \ell_i + \beta \cdot m_i \cdot \frac{1}{\theta(G, \mathcal{M}_i)} \right) \nonumber \\ 
&= s \cdot \alpha + \sum_{i=1}^s \delta\cdot \ell_i + \beta \cdot \sum_{i=1}^{s} m_i \cdot \frac{1}{\theta(G, \mathcal{M}_i)}
\end{align}
\noindent 
\begin{graybox}
\begin{observation}[Collective Completion Time as $\alpha$--$\beta$ Cost]
The classic $\alpha$--$\beta$ cost model for collective communication emerges naturally when we express collective completion time in terms of latency factor $\alpha$, bandwidth factor $\beta$, and importantly, propagation delay $\delta$ and congestion factor which is the inverse of concurrent flow $\theta$. This formulation grounds the cost model in network throughput and reveals its dependence on both the underlying topology and the structure of the collective.
\end{observation}
\end{graybox}

While the $\alpha$--$\beta$ model is widely used in practice, network throughput, propagation delays, and congestion are rarely made explicit in its formulation. A few exceptions relate congestion to communication distance or the number of the messages on a link in structured topologies~\cite{https://doi.org/10.1002/cpe.1206,10.1145/2686882,295653}, but these are often limited to specific patterns or architectures assuming unsplittable flow. On the algorithm synthesis side, Liu et al.~\cite{10.1145/3651890.3672249} recently extended the collective cost model using a multi-commodity flow formulation to capture capacity constraints and routing flexibility for the demand matrix represented by the overall collective operation.
Although prior work has implicitly explored aspects of this connection, our formulation explicitly links the $\alpha$--$\beta$ model to network throughput via concurrent flow. This yields a more comprehensive understanding of performance that accounts for both communication structure and network topology. Notably, our formulation applies to arbitrary topologies, making it broadly applicable beyond structured or hierarchical networks.

\subsection{Optimization Framework for Circuit Switching}
\label{sec:optimization}

The key insight from our observations is that the completion time of a collective communication algorithm is fundamentally tied to the path lengths,  congestion and throughput of the underlying topology in each step. This is especially relevant for circuit switching in photonic interconnects: congestion and path lengths can be reduced to $1$ --- i.e., full throughput --- by establishing direct, high-bandwidth optical paths that exactly match the communication pattern $\mathcal{M}_i$ for each step $i$.

However, realizing these direct paths requires reconfiguring the interconnect, which incurs a reconfiguration delay $\alpha_r$. This creates a clear trade-off: reconfiguring reduces congestion and improves throughput but adds latency, while maintaining a static topology avoids reconfiguration costs but may suffer higher congestion.

This tension opens up an opportunity for optimization: how should we schedule interconnect reconfigurations to minimize the total completion time for any given collective? For example, one might choose to maintain a static topology to avoid reconfiguration overhead but pay persistent congestion costs, or reconfigure before every step to eliminate congestion while incurring the maximum reconfiguration penalty. An effective circuit switching schedule must strike a balance, reconfiguring only in steps when the throughput gain outweighs the cost.

Given any collective communication algorithm with $s$ steps, each with a communication pattern $\mathcal{M}_i$ and data volume $m_i$, we can formulate the following optimization problem. We define two binary variables $x_i$ and $z_i$ as follows:
\begin{equation}
x_i = 
\begin{cases}
1 & \text{base topology}\ G \\
0 & \text{matched topology}\ \mathcal{M}_i\ for\ step\ i
\end{cases}
\end{equation}
\begin{equation}\label{eq:zi}
z_i = 
\begin{cases}
1 & \text{if step $i-1$ and $i$ are both base topologies}\ G \\
0 & \text{otherwise}
\end{cases}
\end{equation}
Here, $x_i$ defines the circuit switching schedule, i.e., whether each step uses the base topology $G$ or a topology that perfectly matches the communication pattern $\mathcal{M}_i$ for step $i$ of the collective. The variable $z_i$ defines whether the interconnect incurs any reconfiguration delay between step $i-1$ and $i$.

The optimization problem can now be formulated as:
\begin{align}
\min & \quad \delta \cdot \sum_{i=1}^s \left(\overbrace{x_i\cdot \ell_i}^{\substack{\text{propagation delay}\\\text{w/o reconf.}}} + \overbrace{(1-x_i)\cdot 1}^{\substack{\text{direct}\\\text{with reconf.}}}\right) \quad + \sum_{1}^{s} \overbrace{(1-z_i)\cdot \alpha_r}^{\substack{\text{reconf. delay}}}  \nonumber \\
&\quad + s\cdot \alpha +\quad \beta \cdot \sum_{i=1}^{s} m_i \cdot (\underbrace{x_i \cdot \frac{1}{\theta(G, \mathcal{M}_i)}}_{\substack{\text{congestion}\\\text{w/o reconf.}}} + \underbrace{(1-x_i)\cdot 1}_{\substack{\text{no congestion}\\ \text{with reconf.}}}) \nonumber \\
& \text{subject to} \quad z_i \ge x_i + x_{i-1} -1 \nonumber \\
& \quad \quad \quad \quad \quad \  z_i \le x_i ;\quad z_i \le x_{i-1} \quad \forall i \in [1, s], \ x_0 = 1 \nonumber \\
& \text{Variables} \quad x_i \in \{0,1\} ; \quad z_i \in \{0,1\}
\end{align}

\begin{figure*}
\centering
\begin{subfigure}{0.238\linewidth}
\centering
\includegraphics[width=\linewidth]{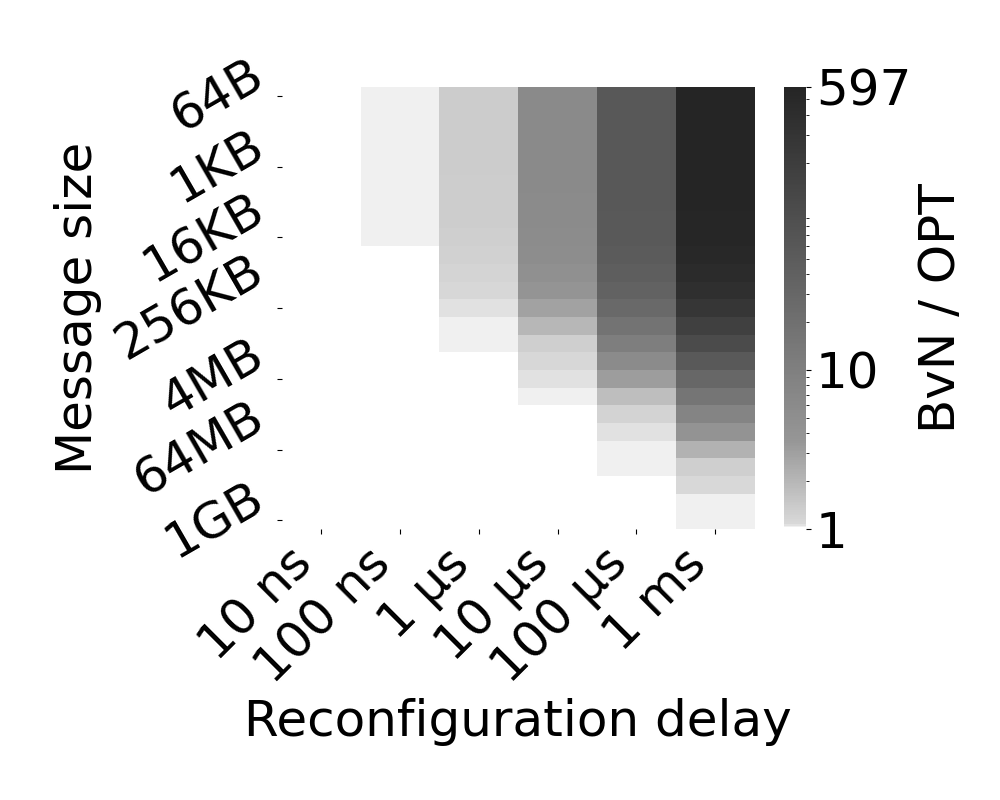}
\caption{\Small{Recursive doubling} \small $\alpha = 100\ ns$}
\label{fig:rd-bvn-100ns}
\end{subfigure}
\begin{subfigure}{0.238\linewidth}
\centering
\includegraphics[width=\linewidth]{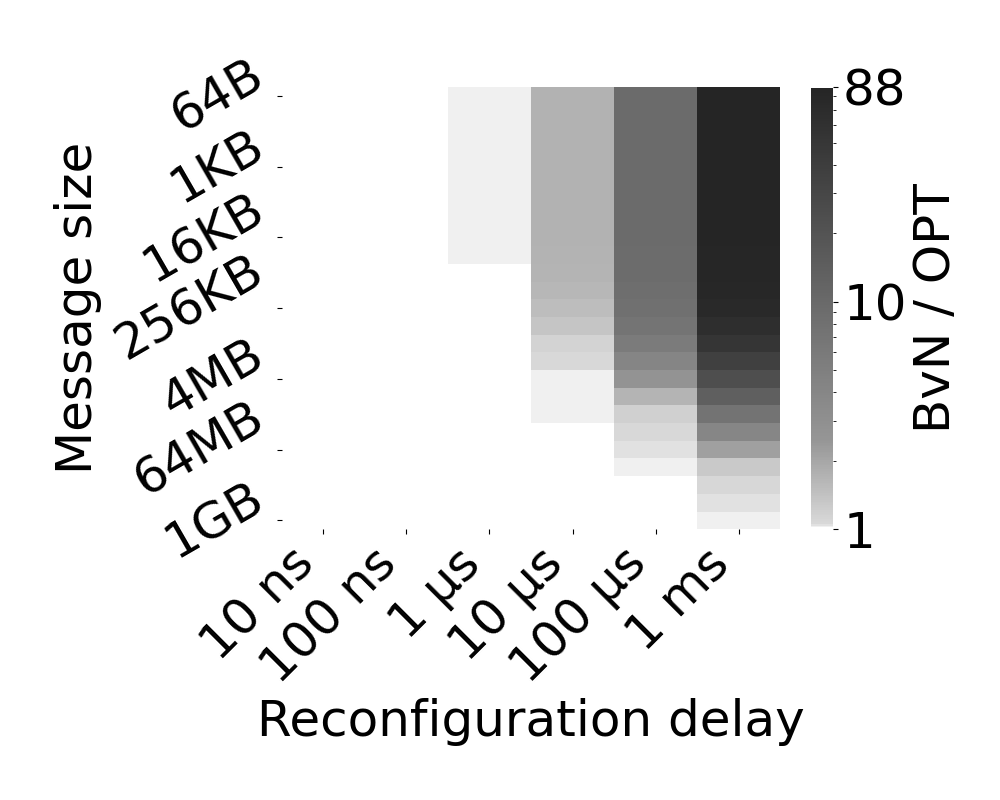}
\caption{\Small{Recursive doubling} \small $\alpha = 10\ \mu s$}
\label{fig:rd-bvn-10us}
\end{subfigure}
\begin{subfigure}{0.238\linewidth}
\centering
\includegraphics[width=\linewidth]{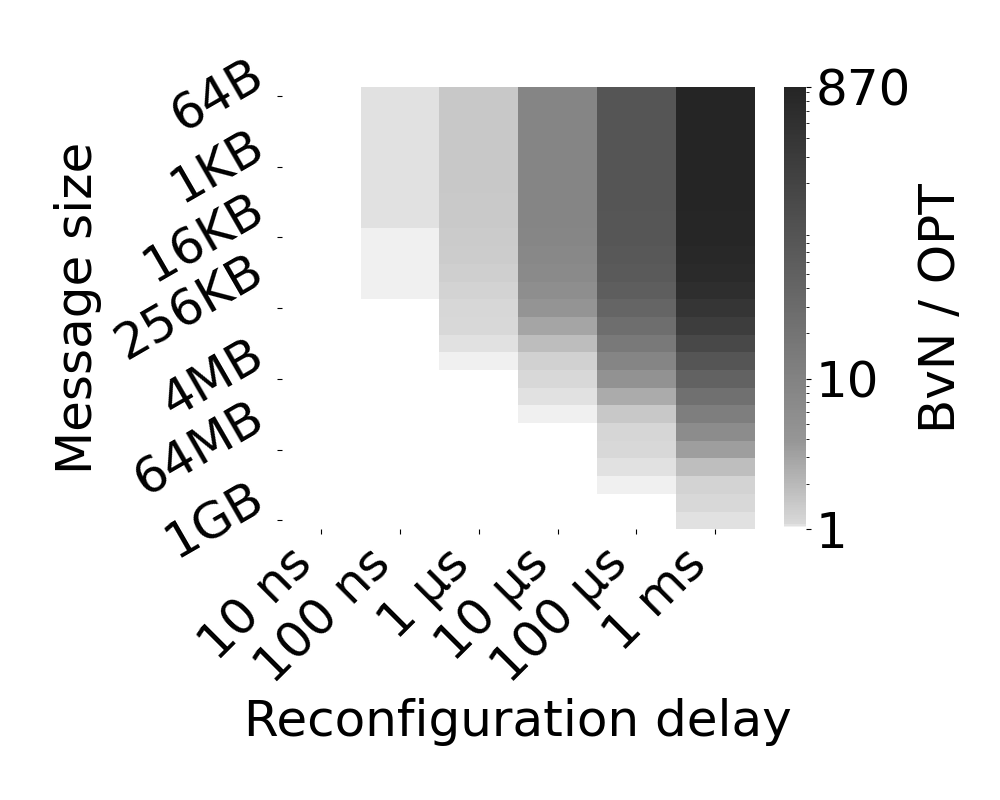}
\caption{\Small{Swing} \small$\alpha = 100\ ns$}
\label{fig:swing-bvn-100ns}
\end{subfigure}
\begin{subfigure}{0.238\linewidth}
\centering
\includegraphics[width=\linewidth]{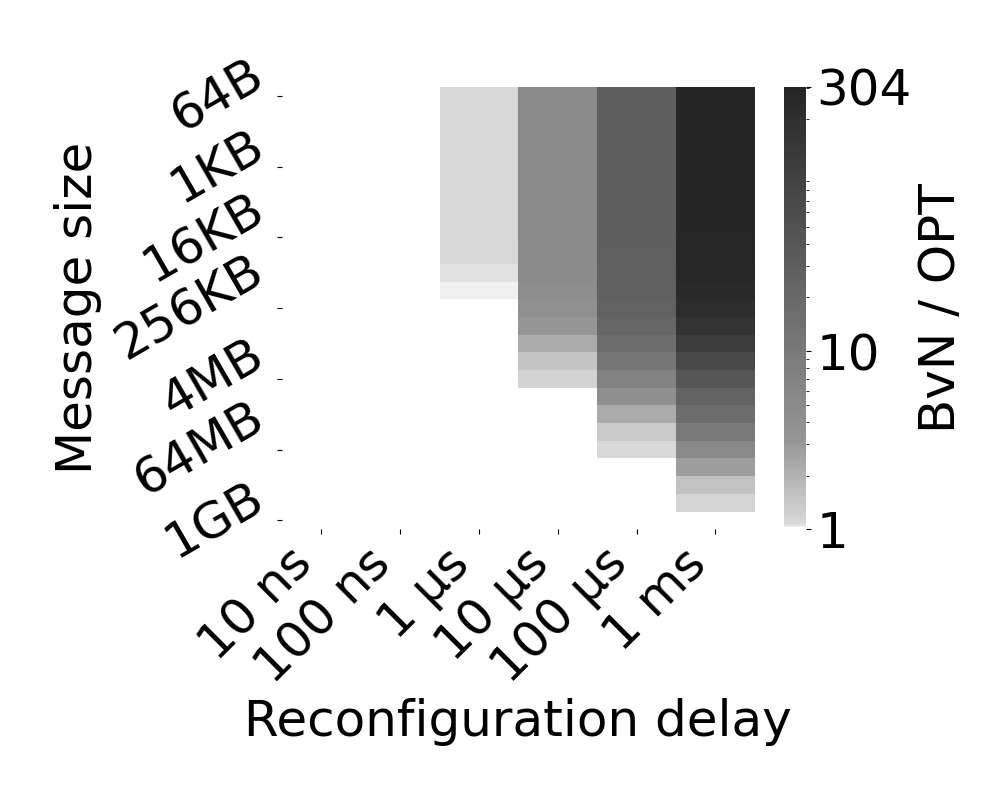}
\caption{\Small{All-to-All} \small$\alpha = 100\ ns$}
\label{fig:swing-bvn-10us}
\end{subfigure}
\begin{subfigure}{0.238\linewidth}
\centering
\includegraphics[width=\linewidth]{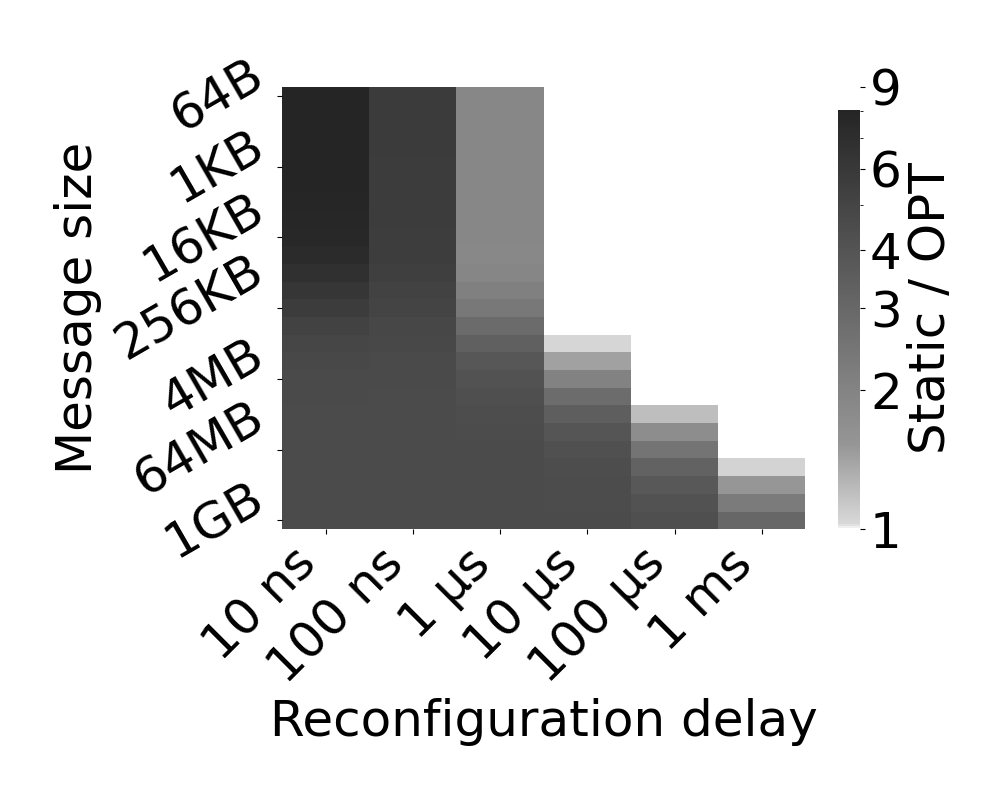}
\caption{\Small{Recursive doubling} \small $\alpha = 100\ ns$}
\label{fig:rd-static-100ns}
\end{subfigure}
\begin{subfigure}{0.238\linewidth}
\centering
\includegraphics[width=\linewidth]{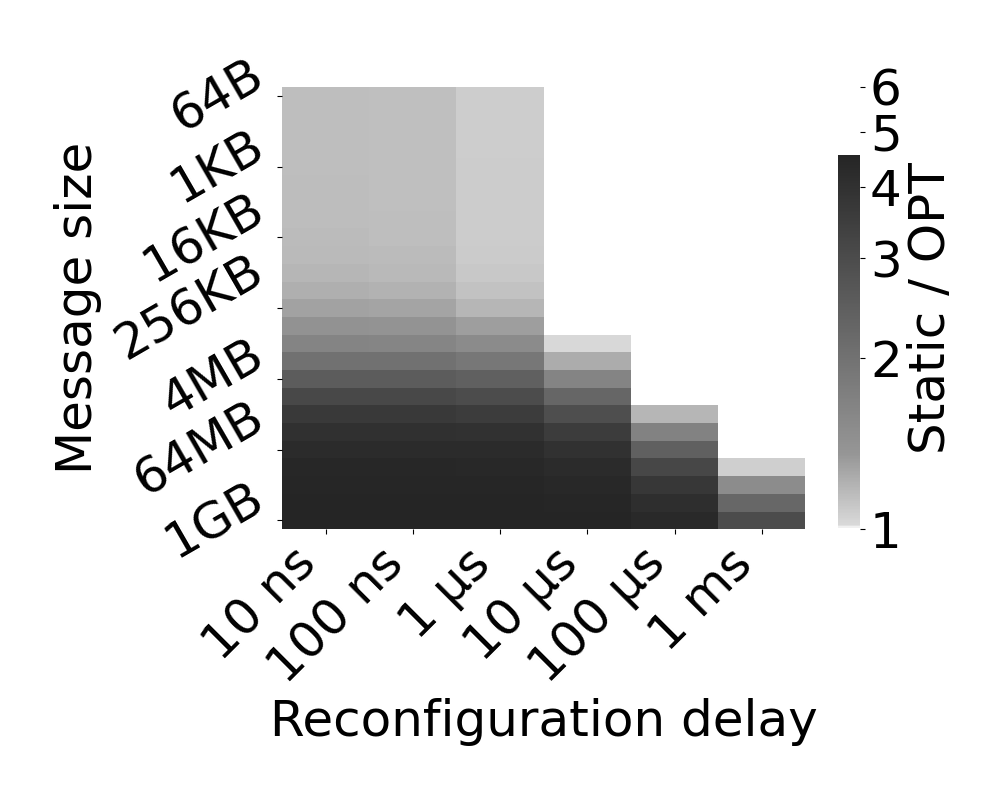}
\caption{\Small{Recursive doubling} \small $\alpha = 10\ \mu s$}
\label{fig:rd-static-10us}
\end{subfigure}
\begin{subfigure}{0.238\linewidth}
\centering
\includegraphics[width=\linewidth]{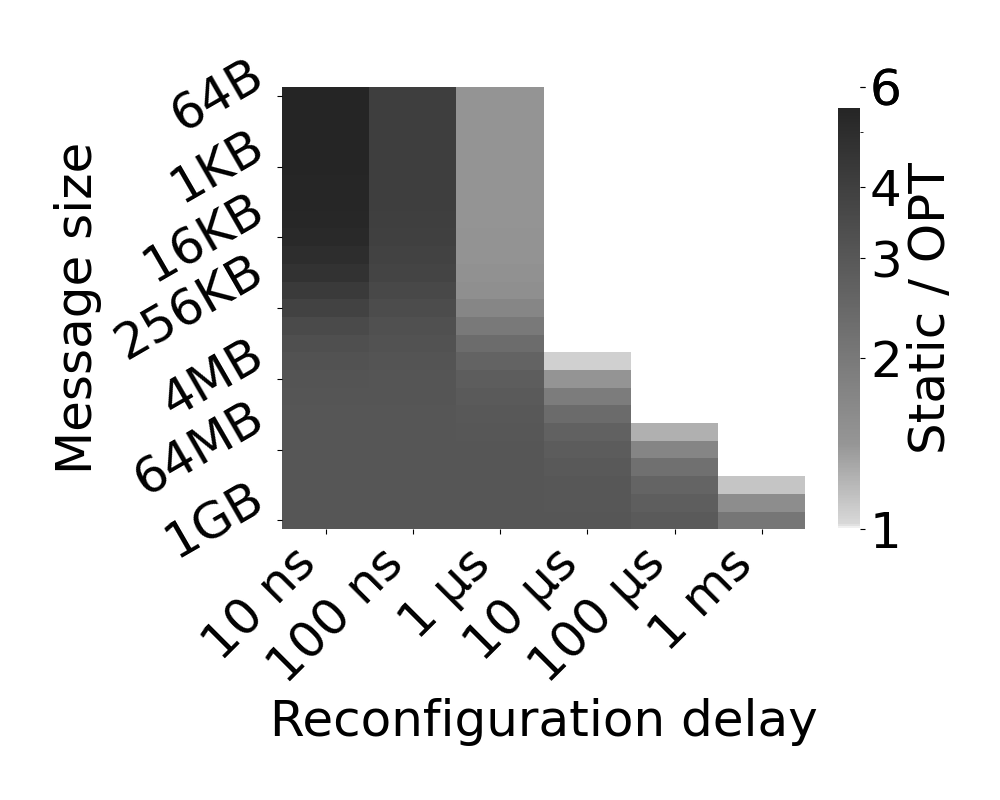}
\caption{\Small{Swing} \small $\alpha = 100\ ns$}
\label{fig:swing-static-100ns}
\end{subfigure}
\begin{subfigure}{0.238\linewidth}
\centering
\includegraphics[width=\linewidth]{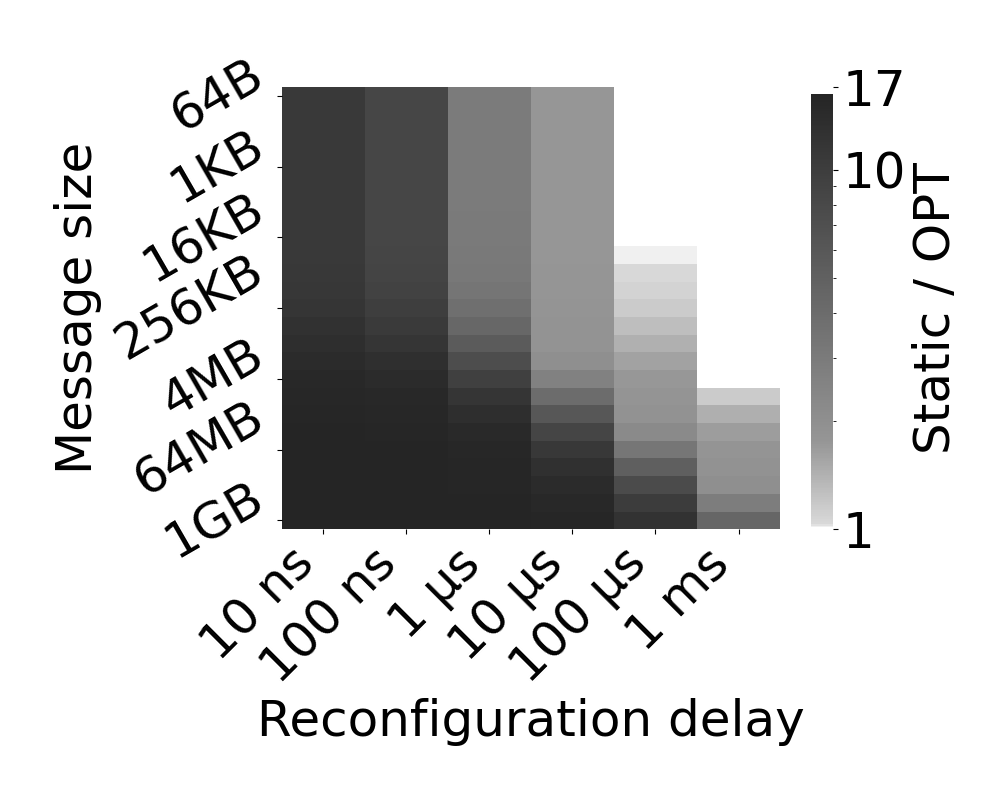}
\caption{\Small{All-to-All} \small$\alpha = 100\ ns$}
\label{fig:swing-static-10us}
\end{subfigure}
\vspace{-4mm}
\caption{Heatmaps showing the speedup in collective completion times achieved by our optimized schedules, compared to BvN-based schedules (top row) and a static ring topology (bottom row).}
\vspace{-4mm}
\label{fig:eval}
\end{figure*}

Our objective is to minimize the total completion time of the collective communication algorithm, which consists of four components: ($\delta\cdot \ell_i$) the propagation delay as function of path lengths, ($\alpha)$ the fixed latency factor, ($\alpha_r$) the total reconfiguration delay incurred by the interconnect, and ($\frac{1}{\theta}$) the congestion factor across all steps. The congestion and propagation delay depend on whether we choose to reconfigure the interconnect to match the communication pattern $\mathcal{M}_i$ or maintain the base topology $G$. The constraints ensure that $z_i$ correctly captures whether a reconfiguration occurs between steps, and all variables are binary.

Overall, this formulation is a mixed integer program ($0$–$1$ ILP), which is NP-hard in the general case~\cite{karp2009reducibility}. Interestingly, our model has a special sequential structure: the variables $x_i$ (interconnect state) and $z_i$ (reconfiguration event) depend only on the previous step. This structure admits an efficient dynamic programming solution and is polynomial-time solvable due to the principle of optimality~\cite{bertsekas2012dynamic}.

This framework captures the fundamental trade-off between reconfiguration delay and congestion in adaptive photonic interconnects. It provides a systematic way to optimize circuit switching schedules for collective communication, balancing the benefits of reconfiguration against its costs. Notably, the optimization is aware of both the data volume in each step and the underlying network throughput. Furthermore, the formulation supports any base topology $G$ and applies to any collective communication algorithm (including custom ones) that can be expressed as a sequence of matchings, or even a sequence of such collective communication operations e.g., All-to-All after an AllReduce operation. 
Our formulation can even be extended to account for a fixed pool of base topologies instead of a single base topology $G$ that we current use \eg using multiple co-prime rings as base topologies or a union of such rings for higher degree networks~\cite{285119}.
Optimizing the base topologies opens further opportunities for performance gains.

\subsection{So What is the $\Delta$ After All? Reconfigure or Not?}
\label{sec:opportunities}

Our focus so far has been on the underlying theoretical problem of optimizing circuit switching for collective communication. But the central question remains: \emph{what performance gain can we actually expect from programmable silicon photonic interconnects?} In other words, for what range of reconfiguration delays does a \emph{programmable} interconnect yield meaningful speedup for collective operations in scale-up domains?

To explore this question, we conduct preliminary evaluations using a flow-level simulator that implements the optimization framework described in \S~\ref{sec:optimization}. We model a scale-up system with $n = 64$ GPUs, each equipped with a single link to a reconfigurable photonic interconnect as introduced in \S\ref{sec:architecture}. We set the link bandwidth to $800$\,Gbps, propagation delay $\delta$ to $100$ns~\cite{295653}, and vary the fixed per-step latency $\alpha$, the reconfiguration delay $\alpha_r$, and the message size. We run the AllReduce collective using recursive doubling and Swing algorithms~\cite{10.1007/978-3-540-24685-5_1,295653} (which are bandwidth-optimal); and All-to-All (transpose) collective. Due to space constraints, we omit the combinations of $\alpha$, $\alpha_r$, and bandwidth, but similar trends hold throughout the full parameter space.
Since each GPU has a single fat link, we use a ring as the base topology $G$ --- a common choice for scale-up photonic interconnects. While our optimization framework is especially valuable for degree $>2$ networks, we use this simple case to clearly illustrate the main trade-offs.
We compare two approaches: (1) a static ring topology, and (2) a reconfigurable interconnect that follows BvN schedules matched to the communication pattern (see \S\ref{sec:concurrent-flow}). We report speedup in terms of the completion time of the collective achieved by our optimized schedules (OPT) compared to these alternatives.

Figure~\ref{fig:eval} summarizes the results. Figures~\ref{fig:rd-bvn-100ns}---\ref{fig:swing-bvn-10us} show the speedup relative to BvN schedules, while Figures~\ref{fig:rd-static-100ns}---\ref{fig:swing-static-10us} show the speedup relative to the static ring. Each column (x-axis) in the heatmaps corresponds to a different value of reconfiguration delay $\alpha_r$, and each row (y-axis) corresponds to a different message size. The color indicates the speedup achieved by our optimized schedule, with darker shades representing higher speedup and no color (or white) indicates speed up of $1$.

\setlength{\intextsep}{2pt}
\begin{wrapfigure}{r}{0.5\linewidth}
  \centering
  \includegraphics[width=\linewidth,clip]{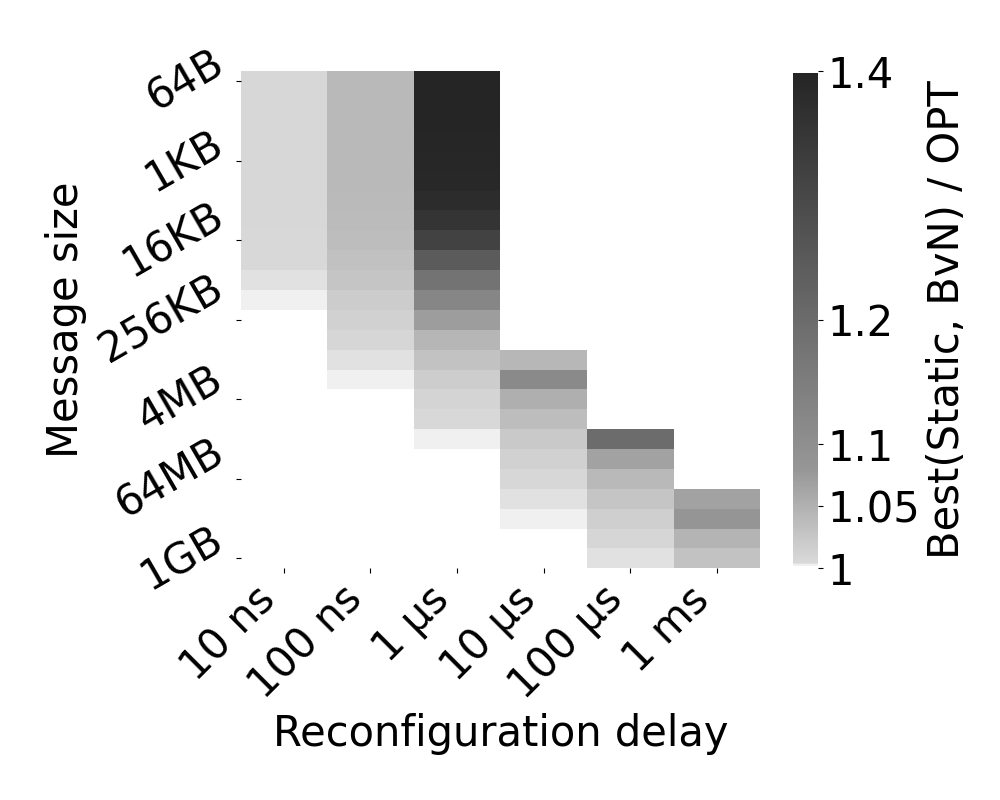}
  \caption{Our optimized schedules can significantly speed up collective communication even compared to the best of both worlds --- BvN schedules and a static ring topology.}
  \label{fig:best}
\end{wrapfigure}

Overall, we see that our framework captures two distinct regimes: significant performance gains (up to orders of magnitude) over BvN schedules appear when reconfiguration delay is high or message sizes are small, where naive per-step reconfiguration would otherwise incur excessive latency. In comparison to a static ring topology, we observe substantial speedup when reconfiguration delay is low and message sizes are large, where our optimized schedule fully exploits the available bandwidth. Interestingly, Figure~\ref{fig:best} shows that there is also a transitional regime --- visible as the diagonal region --- where our optimized schedules outperform both static and naive BvN approaches by adaptively deciding when to reconfigure and when not to. This illustrates precisely \emph{when} adaptive photonic interconnects should reconfigure and when they should not.

\section{Research Agenda and Future Outlook}
\label{sec:vision}

We see many opportunities for performance optimization from a theoretical perspective, along with practical challenges that must be addressed before adaptive photonic interconnects can be fully realized in scale-up domains. We outline a research agenda spanning algorithm design, and systems integration.

\myitem{Fast heuristics for adaptive photonic interconnects:}
As scale-up domains grow, fast heuristics for optimizing circuit-switching schedules become paramount. While our framework offers insight into potential gains, practical implementations need algorithms that adapt quickly to arbitrary collectives. For example, threshold-based heuristics could switch between a static topology and a BvN-based schedule depending on when gains outweigh reconfiguration costs~\cite{circuitml}. Balancing near-optimality with computational efficiency will be crucial for real adoption.

\myitem{Simplifying the congestion factor in the cost model:}
Our framework relies on the maximum concurrent flow $\theta(G, \mathcal{M}_i)$ to capture congestion, but computing this exactly can be expensive, particularly for large topologies. Future work could explore approximations or simpler proxies that retain accuracy but reduce overhead. For example, an upper bound on throughput per permutation pattern based on graph degree could reduce the congestion factor to a function of maximum node degree and the number of communicating GPUs. Such simplifications could make scheduling practical at runtime while preserving useful performance insights.

\myitem{Deeper understanding of the propagation delays:} Our formulation in \S\ref{sec:concurrent-flow}, indicates that the completion time of a collective communication algorithm is influenced by the path lengths and congestion. For AllReduce algorithms, this implies that the ring algorithm is optimal even for short messages if the propagation delays are high. Naturally, recursive doubling~\cite{10.1007/978-3-540-24685-5_1}, or other algorithms like Swing~\cite{295653} that finish in fewer steps become more attractive for reconfigurable interconnects, than for static interconnects. We leave it for future work, to design fast heuristics for AllReduce operations.

\myitem{Routing challenges:}
Reconfigurable interconnects naturally introduce dynamic routing challenges. While a topology that matches a collective step’s pattern allows simple one-hop routing, practical schedules may include intermediate topologies that balance reconfiguration cost against performance. This requires routing algorithms that adapt quickly while maintaining high throughput and low latency. Exploring lightweight, topology-aware routing techniques for dynamic configurations is an important direction.

\myitem{Tackling variable reconfiguration delays:} Our formulation assumes a constant reconfiguration delay $\alpha_r$, but in practice, this may vary with the number of ports or the specific operation. We plan to extend our framework to account for variable delays by modeling them as a function of port count or reconfiguration complexity. This would enable more accurate scheduling that adapts to the interconnect’s characteristics.

\myitem{Overlapping reconfiguration with computation:} Many collectives offer potential to overlap reconfiguration with computation, letting GPUs prepare data while the interconnect reconfigures. We plan to explore how to schedule these overlaps to minimize total completion time, by modeling computation phases as part of the optimization.

\smallskip
Many interesting questions remain open, including extending our model to multi-ported collectives where each step is not a single permutation but a union of multiple permutations; identifying optimal sets of base topologies; and addressing practical aspects such as synchronization.

\smallskip
We envision a future where scale-up GPU systems seamlessly harness the power of reconfigurable photonic fabrics to break through today's bandwidth and energy walls. By bridging theory and practice --- from fast scheduling heuristics to topology-aware routing and reconfiguration-aware collectives --- we can unlock the full potential of adaptive photonic interconnects. Realizing this vision will require close collaboration across systems, networking, and photonics communities, but the payoff is compelling: a new class of datacenter and HPC architectures where communication is entirely in photonic domain, and light truly bends to the collective will.

\label{bodyLastPage}

\bibliographystyle{plain}
\bibliography{references}

\label{LastPage}
\end{document}